\pdfoutput=1

\documentclass[11pt,table]{article}

\usepackage[preprint]{acl}
\usepackage[ruled,vlined]{algorithm2e}
\usepackage{amsmath}
\usepackage{amssymb}
\usepackage{times}
\usepackage{latexsym}
\usepackage{multirow}
\usepackage{cite}
\usepackage{xcolor}
\usepackage{makecell}
\usepackage{subcaption}
\usepackage[normalem]{ulem}
\usepackage{amsthm}
\usepackage{booktabs}
\usepackage{lscape}
\usepackage{breqn}
\usepackage{pifont}%
\usepackage{tabularx}
\usepackage{adjustbox}
\usepackage{array}
\usepackage{pifont}
\usepackage[T1]{fontenc}

\usepackage[utf8]{inputenc}

\usepackage{microtype}

\usepackage{inconsolata}

\usepackage{graphicx}

\usepackage{bm}
\newcommand{\EQ}{\begin{eqnarray}}
\newcommand{\EN}{\end{eqnarray}}

\usepackage{tabulary}

\newcommand*{\rowstyle}[1]{
	\gdef\@rowstyle{\leavevmode#1}%
	\@rowstyle\ignorespaces}

\usepackage{multicol}
\usepackage{algorithmic}
\usepackage{soul, xcolor}
\newcommand{\sysname}{DAMA}

\title{Adapting Where It Matters: Depth-Aware Adaptation for Efficient Multilingual Speech Recognition in Low-Resource Languages}

\author{
 \textbf{Yang Xiao},
  \textbf{Eun-Jung Holden},
 \textbf{Ting Dang}
\\
 The University of Melbourne
\\
 \textsuperscript{}{\small{\textit {yxiao9550@student.unimelb.edu.au}, \textit {\{eunjung.holden, ting.dang\}@unimelb.edu.au}}}
}

\newcommand{\cmark}{\ding{51}}%
\begin{document}
\maketitle
\begin{abstract}

Recent speech foundation models excel at multilingual automatic speech recognition (ASR) for high-resource languages, but adapting them to low-resource languages remains challenging due to data scarcity and efficiency constraints. Full-model fine-tuning is computationally expensive and prone to overfitting, while parameter-efficient methods like LoRA apply adaptation uniformly across layers, overlooking internal representations thus compromising effectiveness and efficiency. We analyze multilingual ASR models and reveal a U-shaped adaptability pattern: early and late layers are language-specific and require more adaptation, while intermediate layers retain shared semantics and need less. Building on this observation, we propose DAMA, a Depth-Aware Model Adaptation framework that allocates adaptation capacity according to each layer’s role. DAMA also introduces Singular Value Decomposition (SVD)-based initialization to constrain adaptation and preserve the U-shaped pattern, as well as a frozen middle-layer basis for further efficiency.
Evaluated on 18 low-resource languages across two benchmark datasets, DAMA matches or surpasses state-of-the-art accuracy with 80\% fewer trainable parameters, achieves a 29\% error reduction under extreme data scarcity, and significantly improves memory, training time, and computational efficiency over baselines. These results highlight the benefits of structure-aware adaptation for efficient, scalable multilingual ASR.


\end{abstract}

%

\section{Introduction}

Recent 
speech foundation models~\citep{mms, cui2025recent} are pretrained on vast amounts of multilingual speech data and capable of performing a variety of tasks, such as multilingual automatic speech recognition (ASR), with particular success in high-resource languages such as English. However, their performance drops substantially on low-resource languages or local dialects~\citep{shi2024ml}, 
due to the limited data availability, resulting in substantial disparities for underrepresented languages.

Efficiently adapting these models for low-resource languages with limited data remains a significant challenge. In many real-world scenarios, such as the rapid deployment of speech technologies in emerging markets or for local dialects, there is a critical need for adaptation methods that are both accurate and highly efficient. These approaches must enable fast adaptation to limited new data, operate within stringent memory and computational constraints, and maintain strong performance even in extremely low-data settings.

Traditional full-parameter fine-tuning is computationally and memory intensive, making it impractical for many applications, especially on edge devices or with limited resources. The scarcity of labeled data for low-resource languages further complicates adaptation, as full-parameter fine-tuning often suffer from overfitting or catastrophic forgetting~\citep{wang2024comprehensive,chang2021towards,yang2022online,kwok2024continual}.


Recent work has shifted towards parameter-efficient fine-tuning (PEFT)~\citep{ding2023parameter}, such as Low-Rank Adaptation (LoRA)~\citep{hu2022lora}, which freezes the original model parameters and updates only a small set of new weights for the target language. These methods enable faster and more memory-efficient adaptation. However, these methods adapt all layers uniformly in a brute-force manner, overlooking the structure of language representations and potentially limiting effectiveness and parameter efficiency, especially in extremely low-resource settings. In large models, even standard LoRA may still require updating a significant number of parameters.

Therefore, advancing efficient multilingual adaptation in speech foundation models requires a nuanced understanding of how these models represent and share multilingual knowledge internally. This study first conducts a layer-wise analysis of how multilingual speech representations are maintained and interact across different model layers, revealing a distinct U-shaped pattern of plasticity: early and late layers capture language-specific features and are more adaptable, while middle layers remain language-agnostic. This suggests that different layers require varying degrees of adaptation to new languages, challenging the prevailing assumption in prior work that all layers are equally suitable for adaptation~\citep{song2024lora,kwok2025two}.

Motivated by this U-shaped pattern, we propose Depth-Aware Model Adaptation (DAMA), which introduces three mechanisms to tune models for new languages with both effectiveness and efficiency. First, the Depth-Aware Rank Schedule allocates higher adaptation capacity to the more plastic early and late layers while restricting the rank in the middle layers, balancing parameter efficiency with preservation of the model structural properties. Second, to constrain adaptation in the middle layers, we propose SVD-Based Initialization, which initializes adaptation weights in directions orthogonal to the dominant weights of the model. This helps preserve shared language representations and maintain the U-shaped adaptability. Finally, to further improve efficiency, especially in low-resource settings, we introduce Basis-Protected Projection (BPP), where a subset of adaptation weights is frozen, thus reducing the number of trainable parameters while preserving essential knowledge.

We evaluated \sysname{} on 18 low-resource languages using the Common Voice and FLEURS datasets. \sysname{} matches or outperforms state-of-the-art baselines while reducing parameters by about 80\%. More importantly, in extremely low-resource settings (0.5 to 1 hour of data), it achieves up to 29\% relative Word Error Rate (WER) improvement. Efficiency analysis shows a 24\% gain in GPU memory utilization and 36\% faster training. These results highlight that adaptation aligned with model layer properties enables scalable, parameter-efficient multilingual systems without sacrificing accuracy.
Our contribution is summarized below:
\begin{itemize}
    
\item  We are the first to systematically analyze layer wise multilingual language representations in speech foundation models. We reveal a U-shaped distribution of language specificity, which demonstrates how these models maintain and share cross-lingual knowledge.

\vspace{-1mm}
    \item We propose DAMA, a novel depth-aware multilingual ASR adaptation framework that achieves an effective balance between adaptability and parameter efficiency, while preserving essential multilingual knowledge.
\vspace{-1mm}
\item Our experiments on 18 languages demonstrate a Pareto-optimal trade-off for the proposed \sysname{}. It exceeds or matches SOTA performance while significantly reducing trainable parameters, memory usage and training time. Most importantly, DAMA exhibits superior robustness in low-resource settings. 
\end{itemize}

\section{Related Work}
\paragraph{Multilingual Automatic Speech Recognition (MASR).} MASR ~\citep{yadav2022survey} aims to transcribe speech across diverse languages using a single, unified foundation model. Recent advancements have been driven by scaling up training data and model capacity, such as Whisper~\citep{whisper} and MMS~\citep{mms}. 
While these models excel at recognizing high-resource languages, their performance degrades significantly when applied to low-resource languages unseen during pre-training. 
Due to the limited data availability, it  
necessitates the efficient adaptation to new languages without the prohibitive computational cost of full retraining or the risk of catastrophic forgetting~\citep{li2022massively}. 
\paragraph{Efficient Multilingual Adaptation Strategies.} To adapt  speech foundation models to new languages, fully fine-tuning has traditionally been the default approach. However, updating all parameters is computationally prohibitive for resource-constrained settings. More importantly, FFT often leads to catastrophic overfitting where the model memorizes sparse data at the expense of generalizable semantic knowledge~\citep{cl2,yang2022online,peng2024dark}. Some methods attempt to preserve prior knowledge by constraining updates to important parameters~\citep{xiao2025analytickws,xiao25c_interspeech}. However, these methods often struggle to balance the plasticity-stability dilemma, limiting their ability to learn new tasks effectively. To enable efficient adaptation, PEFT has become the main paradigm.  Methods such as Low-Rank Adaptation (LoRA)~\citep{hu2022lora} and Adapters~\citep{houlsby2019parameter} freeze the pre-trained backbone and inject a small number of trainable parameters to capture task-specific shifts. Despite their success, standard PEFT methods in MASR overlook language representations and adapt all model layers uniformly, compromising both efficiency and effectiveness~\citep{song2024lora,li25p_interspeech,yang25m_interspeech}. 
While methods like AdaLoRA~\citep{zhangadaptive} introduce dynamic rank allocation, they depend on sensitivity scores derived from training data. In few-shot or low-resource settings, insufficient data renders these sensitivity estimates unstable
because they rely on computationally expensive searches. 
This highlights a critical gap in more effective and efficient adaptation mechanisms for low-resource speech data. 
\\

\section{
Analysis of Language Representations}
To investigate how language representations are maintained and shared within the latent space, we conduct a layer-wise analysis utilizing linear probing at each layer to perform a language identification (LID) task. If the representations at a given layer are language-specific, the LID accuracy will be high; conversely, lower accuracy suggests more language-agnostic representations. This analysis enables us to identify the extent to which language-specific information is preserved across layers.

\subsection{Layer-wise Linear Probing }Specifically, we conduct our analysis within the encoder-decoder framework, which is commonly used in recent speech foundation models~\citep{whisper,oasr,owsm}. Given a speech input sequence \( x_t \), the encoder processes this input and generates a sequence of latent representations, denoted as \( \mathbf{h}_t = \mathrm{Encoder}(x_t) \). These encoder outputs \( \mathbf{h}_t \) are then provided as input to the decoder, \( f_\theta \), parameterized by weights \( \theta \). The decoder consists of \( L \) layers, with the intermediate activations at the \( l \)-th layer denoted as \( \mathbf{z}_t^{(l)} \), where \( l \in \{1, \dots, L\} \).

For each decoder layer \( l \), we apply linear probing to 
the representations \( \mathbf{z}_t^{(l)} \). Specifically, we train a linear classifier \( g^{(l)}(\cdot) \) on the activations \( \mathbf{z}_t^{(l)} \) to perform the LID task with cross-entropy loss. The classification accuracy provides a quantitative measure of the degree to which language-specific information is encoded at the \( l \)-th layer. This provides a comprehensive view of how language representations are maintained, diminished, or abstracted at different depths of the decoder network.


\paragraph{Dataset and Language Selection:} To ensure the robustness of our analysis, we select a diverse set of languages from the Common Voice~\citep{ardila2020common} dataset. We construct two distinct evaluation groups to test the generalization capability of the model. The first group consists of five languages that were seen during the pre-training, which allows us to measure how the model represents known knowledge. The second group consists of ten unseen languages, which allows us to observe how the model handles completely new linguistic patterns. We provide the full list of these languages and their specific details in Appendix A. \\

\begin{figure}[t!]
    \centering
    \begin{subfigure}[t]{\linewidth}
        \centering
        \includegraphics[width=\linewidth]{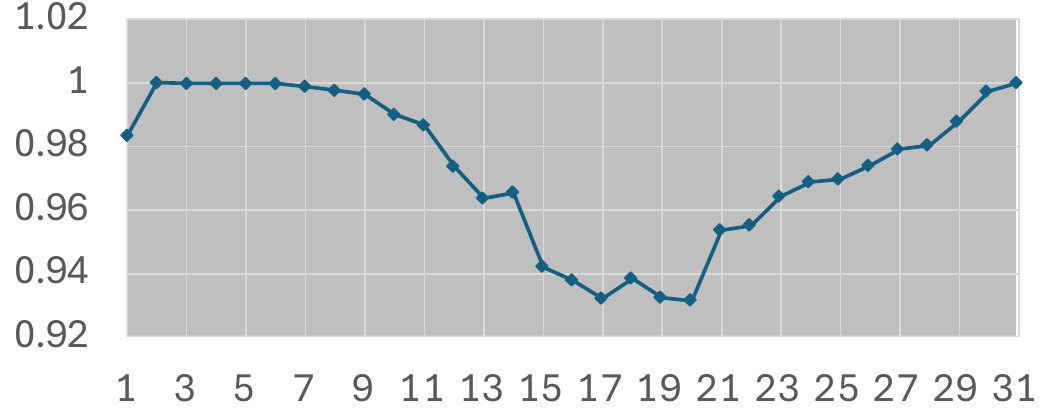}
        \caption{Linear Probing Accuracy of Five Seen Languages.}
        \label{fig:track1}
    \end{subfigure}

    \vspace{4mm} 
    \begin{subfigure}[t]{\linewidth}
        \centering
        \includegraphics[width=\linewidth]{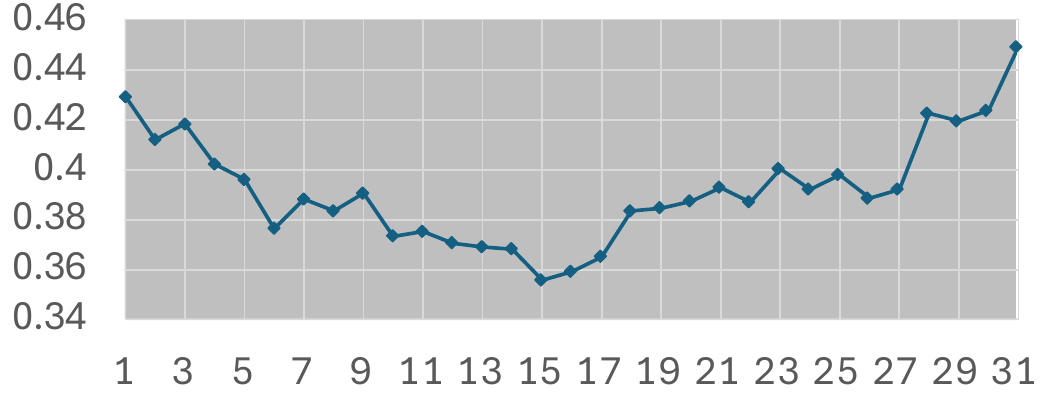}
        \caption{Linear Probing Accuracy  of Ten Unseen Languages.}
        \vspace{4mm}
        \label{fig:track2}
    \end{subfigure}
    \begin{subfigure}[t]{\linewidth}
        \centering
        \includegraphics[width=\linewidth]{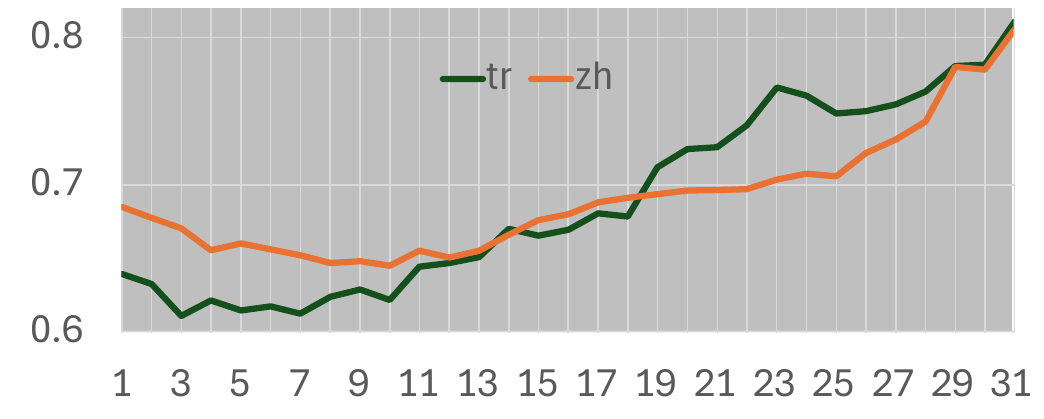}

                \caption{Linear Probing Accuracy of Five Seen Languages after Fine-tuning.}
        \label{fig:track3}
    \end{subfigure}

    \vspace{4mm} 
    \begin{subfigure}[t]{\linewidth}
        \centering
        \includegraphics[width=\linewidth]{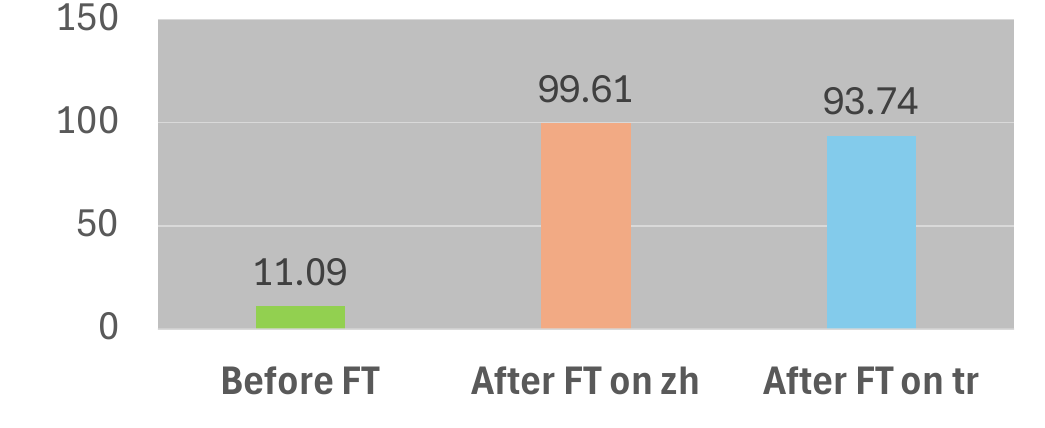}
        \caption{WER after fine-tuning on English.}

        \label{fig:barchart}
    \end{subfigure}

    
    \caption{Layer-wise probing for different languages before and after fine-tuning.
    }
    \label{fig:tracks}
\end{figure}

\subsection{The U-Shaped Layer-wise Plasticity}

\paragraph{Analysis of Seen Languages:} 
The results on seen languages, as illustrated in Figure~\ref{fig:track1}, reveal a distinct U-shaped pattern across the depth of the decoder. The early layers 
(layers 1 to 5) and the late layers 
(layers 28 to 32) achieve a near 100\%. This indicates that the model retains strong language-specific markers. 
Conversely, the intermediate layers exhibit a noticeable drop in performance. Specifically, the accuracy falls to approximately 93\% around layer 17. 
The relative decline forms a ``\emph{Semantic Valley}'', which suggests that the middle layers are less sensitive to language identity and focus more on language-agnostic semantic representations. \\ 

\paragraph{Analysis of Unseen Languages:}
Further analysis of languages that were not seen during pre-training also reveals the same \emph{``Semantic Valley}'', despite the decreased accuracy. 
This consistent U-shaped behavior confirms that decoder naturally organizes information with depth-dependent plasticity.
Regardless of the languages, the early and late layers capture language-specific linguistics features, while the middle layers relatively maintain a language-agnostic representation.

It is interesting to note that this U-shaped pattern aligns with recent findings in text-based Large Language Models~\citep{ushape1,ushape2,ushape3}. Our results demonstrate that this hierarchical phenomenon is also evident in speech foundation models, even with the different end-to-end learning paradigm of speech training. 

\subsection{The Impact of Fine-tuning}

We further investigate how the commonly used fine-tuning affects the U-shaped plasticity. We first perform full parameter fine-tuning on two distinct languages, Turkish (tr) and Mandarin (zh), and repeat the probing analysis. The results, presented in Figure~\ref{fig:track3}, shows that the distinct U-shaped pattern has completely disappeared, where the ``\emph{Semantic Valley}'' observed in the pre-trained model is disrupted by a continuous upward trend. 
The middle layers, which captures language-agnostic representations, have been forced to encode strong language-specific information. While it enables adaptation to new languages, the semantic collapse coincides with catastrophic forgetting of seen languages such as English, as shown in Figure~\ref{fig:barchart}, where the WER on English surges from 11\% to over 93\% after fine-tuning. Further, it also results in significant computational costs, including increased memory footprint and longer training times. 

While standard LoRA improves adaptation efficiency and preserves the original model weights, it does not account for the specialized role of the middle layers. By uniformly adapting all layers, including those responsible for language-agnostic processing, LoRA may still disrupt critical semantic representations, eroding well-established structures and undermining cross-lingual generalization and robustness. Preserving the U-shaped plasticity, particularly in the middle layers, is essential to safeguard learned knowledge and improve efficiency, as adaptation should primarily target the early and late layers.

\begin{figure*}[t!]
\centering  
\includegraphics[width=0.99\linewidth]{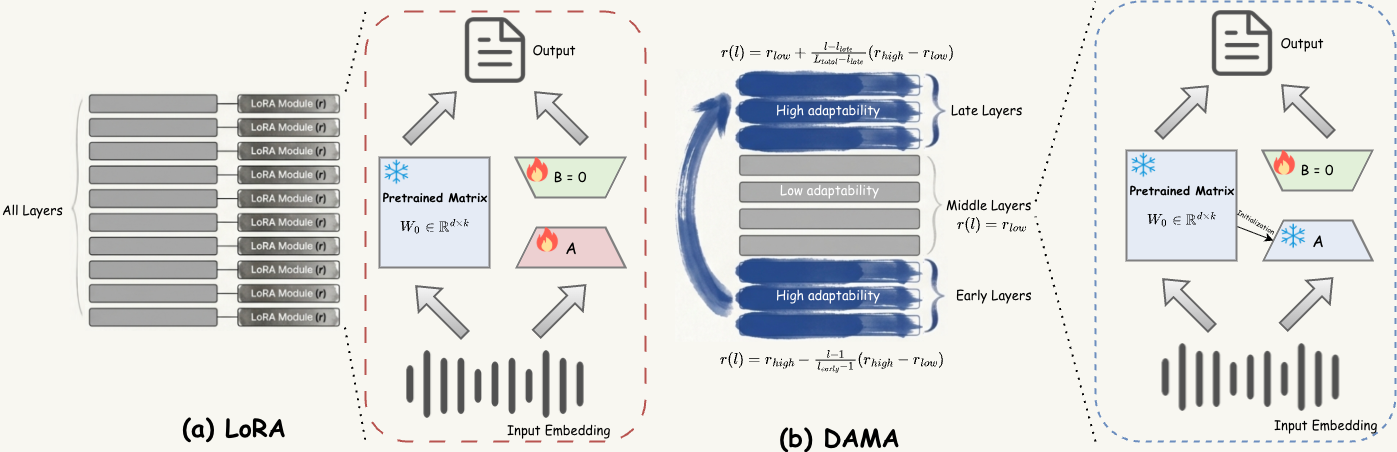}
\caption{Overview of the DAMA Framework compared with LoRA. (a) The standard LoRA with uniform rank. (b) The DAMA Framework. Specifically, the Depth-Aware Rank schedule allocates high plasticity to the early and late layers, while the Basis-Protected Projection physically locks the middle layers to protect ``Semantic Valley''. All layers mean all the layers from the decoder. The decoder processes acoustic embeddings from the encoder and transcribes them into output tokens.
}
\label{fg1}
\vspace{-5mm}
\end{figure*}

\section{Proposed DAMA}


To maintain the U-shaped structure during adaptation, we propose the efficient Depth-Aware Model Adaptation (DAMA). First, We introduce a Depth-Aware Rank Schedule assigning higher adaptation capacity to the early and late layers while restricting the lower adaptation in the middle layers. Second, we propose SVD-based Initialization for LoRA in the middle layers to explicitly preserve language-agnostic representations, enabling efficient low-rank adaptation. Third, we design Basis-Protected Projection (BPP), which protects and freezes parameters in the middle layers to further improve adaptation efficiency and stability, especially in low-resource settings.

\subsection{Depth-Aware Rank Allocation}


Standard LoRA~\citep{hu2022lora} adapts pre-trained models by fine-tuning a low-rank update to each weight matrix. Given a pre-trained model with weight matrix $W_0 \in \mathbb{R}^{d \times k}$, LoRA constrains the update $\Delta W$ as the product of two low-rank matrices: $B \in \mathbb{R}^{d \times r}$ and $A \in \mathbb{R}^{r \times k}$, where $r \ll d, k$. The forward pass is then:
\begin{equation}\label{eq:1}
    h = (W_0 + \Delta W) x = W_0 x + BA x,
\end{equation}
During training, $W_0$ is frozen. $A$ is randomly initialized and $B$ is set to zero, so initially, the update has no effect. Standard LoRA assigns the same rank $r$ to all layers, allocating uniform adaptation capacity to early, middle, and late layers.

Instead, DAMA employs the \textit{Depth-Aware Rank} scheduling following the U-shaped distribution, with 
a layer-dependent rank function \( r(l) \) that flexibly adjusts the adaptation capacity based on the position \( l \) of each decoder layer. Specifically, we divide the $L$ decoder layers into three segments:
\textit{Early} ($l_{\text{early}}: 1 \leq l \leq l_{\mathrm{early}}$),
\textit{Mid} ($l_{\text{mid}}: l_{\mathrm{early}} < l < l_{\mathrm{late}}$), and
\textit{Late} ($l_{\text{late}}: l_{\mathrm{late}} \leq l \leq L_{\mathrm{total}}$),
where $l_{\mathrm{early}} = \lfloor \theta_1 L_{\mathrm{total}} \rfloor$
and $l_{\mathrm{late}} = \lfloor \theta_2 L_{\mathrm{total}} \rfloor$ for $0 < \theta_1 < \theta_2 < 1$.
We assign each layer a LoRA rank $r(l)$ with a U-shaped schedule bounded by a maximum rank \( r_{\text{high}} \) and a minimum rank \( r_{\text{low}} \): 
\begin{align}
r(l_{\text{early}}) &= r_{\text{high}} - \frac{l-1}{l_{\text{early}}-1} (r_{\text{high}} - r_{\text{low}}), \\[0.8ex]
r(l_{\text{mid}})   &= r_{\text{low}}, \\[0.8ex]
r(l_{\text{late}})  &= r_{\text{low}} + \frac{l - l_{\text{late}}}{L_{\text{total}} - l_{\text{late}}} (r_{\text{high}} - r_{\text{low}}),
\end{align}

As shown in Figure 2, this schedule allocates more adaptation capacity to the early layers where the model needs to adapt to the specific characteristics of new target languages, and late layers where the model transitions from
processing semantic representations to generating specific lexical outputs in the target language, while minimizing adaptation in the middle to pre-
serve the language-agnostic space. This improves efficiency and stability across all projection modules.

\subsection{SVD-based Initialization}
While a low rank $r$ in the middle layers helps preserve the characteristic U-shaped representational structure, we further reinforce this by employing SVD-based initialization for adaptation. This approach constrains LoRA updates to directions that minimally impact the language-agnostic semantic subspace, thereby preventing over-adaptation and semantic drift in these critical layers. Unlike standard LoRA, which uses random Gaussian initialization and may introduce "geometric noise," SVD-based initialization helps maintain the integrity of the U-shaped representation.


Given a weight matrix $W \in \mathbb{R}^{m \times n}$ from the middle layers of the pre-trained speech foundation model, we perform SVD:
\begin{equation}
    W = U\, \Sigma\, V^T,
\end{equation}
where $\Sigma = \mathrm{diag}(\sigma_1, \dots, \sigma_{\min(m, n)})$ contains the singular values in descending order. The leading $r$ principal components, corresponding to the top singular values $\sigma_1, \dots, \sigma_r$, capture the language-agnostic semantic knowledge that should be preserved during adaptation.

To constrain the adaptation to directions orthogonal to this subspace, we select the residual components by taking the trailing singular vectors $V_{\mathrm{tail}} = [v_{r+1}, \dots, v_{\min(m, n)}]$. We then initialize the LoRA adaptation matrix $A$ (see Eq.~\eqref{eq:1}) as:
\begin{equation}
    A = V_{\mathrm{tail}}^T,
\end{equation}
This initialization restricts the LoRA updates to directions with minimal overlap with the language-agnostic semantic subspace. As a result, the core semantic structure of the pre-trained model is preserved, reducing the risk of semantic drift during adaptation.

\subsection{Basis-Protected Projection (BPP)}
To further stabilize adaptation and improve computational efficiency, we introduce the BPP module, which freezes the LoRA matrix \(A\) in the middle layers and updates only \(B\). By freezing the SVD-initialized \(A\), we ensure that training updates cannot inadvertently steer the adaptation back into the protected semantic subspace, thereby strictly preserving the core language-agnostic representations and making semantic drift virtually impossible. Unlike standard LoRA, which updates both \(A\) and \(B\), our approach constrains adaptation to a significantly lower-rank space by updating only \(B\). This strategy is particularly advantageous in extremely low-resource settings, as it substantially reduces the number of trainable parameters. Consequently, the risk of overfitting is lowered and generalization may be improved, especially when the downstream task dataset is limited in size.

\begin{table*}[t!]
\centering
\small
\setlength{\tabcolsep}{4pt}
\renewcommand{\arraystretch}{1.15}
\caption{Performance comparison using Average WER and parameter efficiency on Common Voice and FLEURS. The rightmost column separately reports additional MACs for each method.}
\vspace{-3mm}
\resizebox{0.91\linewidth}{!}{%
\begin{tabular}{l c c c c c | c}
\toprule
\toprule
\multirow{2}{*}{Method} & \multirow{2}{*}{Params (M)} 
& \multicolumn{2}{c}{Unseen Languages} 
& \multicolumn{1}{c}{Seen-Weak} 
& \multirow{2}{*}{Average} 
& \multirow{2}{*}{Extra MACs (G)} \\
& & Common Voice & FLEURS & FLEURS & & \\
\midrule
Fine-tuning         & 906.5 & 43.87 & 50.05 & 27.55 & 41.34 & -     \\
LoRA~\citep{hu2022lora}        & 68.2  & 43.25 & 47.13 & \textbf{25.19} & 39.71 & 102.2  \\
DoRA~\citep{dora}              & 68.7  & 43.61 & \textbf{46.46} & 25.24 & 39.73 & 1415.4 \\
LoRA-FA~\citep{lorafa}         & 34.1  & 46.00 & 48.56 & 25.65 & 41.55 & 51.1   \\
LoRA-XS~\citep{loraxs}         & \textbf{1.3} & 57.36 & 55.23 & 30.29 & 50.06 & 104.2  \\
VB-LoRA~\citep{li2024vb}       & 4.9   & 59.81 & 50.75 & 28.00 & 49.59 & 758.3  \\
AdaLoRA~\citep{zhangadaptive}  & 51.1  & 51.66 & 52.36 & 28.41 & 46.02 & 76.7   \\
DAMA (Ours)                    & 14.9  & \textbf{43.20} & 47.25 & 25.26 & 39.73 & \textbf{22.3}  \\
\bottomrule
\bottomrule
\end{tabular}%
\vspace{-20mm}
}
\end{table*}

\section{Experimental Setup}

\subsection{Datasets}

We evaluate our approach using two multilingual datasets to ensure broad applicability: Common Voice~\citep{ardila2020common} and FLEURS dataset~\citep{conneau2023fleurs}. 
Common Voice is a crowd-sourced collection providing validated transcriptions for a wide variety of languages. Following established protocols~\citep{cl-masr,kwok2025two}, we select ten languages that are unseen during training, including \emph{Kinyarwanda, Esperanto, Kabyle, Luganda, Meadow Mari, Central Kurdish, Abkhaz, Kurmanji Kurdish, Frisian, and Interlingua}. For these languages, we adopt a standard data split of ten hours for training, one hour for validation, and one hour for testing.

FLEURS dataset~\citep{conneau2023fleurs} 
is derived from the FLoRes machine translation benchmark and contains parallel speech and text. From FLEURS, we construct two distinct evaluation groups. The first group consists of "Seen Weak" languages, such as \emph{Hindi and Welsh}, which the model has encountered but still struggles to process effectively. The second group consists of ``Unseen'' languages, including \emph{Ganda and Sorani Kurdish}, which are completely new to the model. 

To ensure a fair and rigorous evaluation, we selected these languages based on three strict criteria. 
First, we maximized geographic diversity by including languages from Africa, Europe, and Asia. This prevents regional bias in our results. Second, we prioritized languages with limited resources. These languages often lack sufficient training data and pose a greater challenge than widely spoken languages like English. Finally, we selected languages based on task difficulty. Notably, the baseline Whisper model fails to transcribe the ``Unseen'' group entirely, and can not perform well in the ``Seen Weak'' group. 
For the details of the dataset, please refer to Appendix B. 

\subsection{Implementation Details}
We employ the Whisper large v2 model~\citep{whisper} as the backbone which utilizes a standard encoder-decoder architecture and is a commonly used multilingual speech foundation model. 
Following prior benchmark~\citep{cl-masr}, we initialize and train new token embeddings for unseen languages to ensure the model can identify them correctly. To assess performance, we compare our method against fine-tuning and SOTA PEFT methods, including standard LoRA~\citep{hu2022lora} and its variants: DoRA~\citep{dora}, LoRA-FA~\citep{lorafa}, LoRA-XS~\citep{loraxs}, VB-LoRA~\citep{li2024vb}, and AdaLoRA~\citep{zhangadaptive}. We selected these methods to cover a wide range of efficiency strategies, such as dynamic rank allocation and weight decomposition.

For our method, we apply it to all linear projection matrices, including the Query, Key, Value, Output, and the FFN layers. 
We optimize the \(r_{\text{high}}\) and alpha to 32 while \(r_{\text{low}}\) to 8. We set \(\theta_1\) and \(\theta_2\) to 0.3 and 0.7. Training proceeds for two epochs with a batch size of 6. We use the AdamW optimizer combined with a dynamic learning rate scheduler. Finally, we apply a greedy decoding strategy for all inference tasks. We assess the systems using both accuracy metric of WER and efficiency metrics including number of parameters, MACs (Multiply Accumulate operations) and GPU memory. The MACs specifically measure the additional floating-point operations introduced by the trainable adapter modules, excluding the frozen backbone. Complete configuration details and baseline settings can be referred to Appendix B.

\section{Results and Discussion}

\subsection{Performance Comparison}

Experiments on the Common Voice and FLEURS datasets (Table 1) show that DAMA achieves a superior trade off between accuracy and efficiency. DAMA attains an average WER of 39.73\%, which effectively matches the strongest baseline, LoRA, at 39.71\%. Importantly, DAMA exhibits stronger robustness across different languages: it successfully avoids the significant performance drop seen with compression approaches like VB-LoRA on "Unseen" languages (43.20\% vs. 59.81\%), and it outperforms full fine-tuning on "Seen-Weak" languages (47.25\% vs. 50.05\%). However, DAMA uses only 14.9 million parameters, around one-fifth as many as standard LoRA, which requires 68.2 million parameters. In contrast, full fine-tuning yields a significantly higher WER of 41.34\% despite using over 900 million parameters, more than 60 times than ours. This suggests that updating all parameters is inefficient and likely degrades the source model with limited data. 



\subsection{Data Efficiency in Low-Resource Settings} 
\begin{table}[t!]
\centering
\small
\setlength{\tabcolsep}{6pt}
\renewcommand{\arraystretch}{1.1}
\caption{Avg WER on Common Voice (10 languages) in the 2-hour, 1-hour, and 0.5-hour low-resource setting.}
\label{tab:cv_1h_10langs_wer}
\resizebox{\linewidth}{!}{%
\begin{tabular}{l c c c}
\toprule
Method & Avg WER (0.5h) & Avg WER (1h) & Avg WER (2h) \\
\midrule
Fine-tuning              & 68.11 & 61.79  & 54.60\\
LoRA~\citep{hu2022lora}  & 64.84 & 59.96  & 54.28\\
DoRA~\citep{dora}        & 64.73 & 59.93  &  54.24\\
LoRA-FA~\citep{lorafa}   & 68.63 & 63.23  & 57.48\\
LoRA-XS~\citep{loraxs}   & 80.57                  & 73.99  & 67.58\\
VB-LoRA~\citep{li2024vb} & 73.05 & 76.64 & 77.81\\
AdaLoRA~\citep{zhangadaptive} & 90.16 & 72.73 & 65.94\\
DAMA (Ours)              & \textbf{64.11} & \textbf{58.80}  & \textbf{54.20}\\
\bottomrule
\end{tabular} }
\vspace{-5mm}
\end{table}

We further investigate the Average WER of DAMA in data scarce scenarios, with 0.5 hours to 2 hours of per unseen languages from Common Voice, as shown in Table 2. DAMA maintains remarkable generalization capabilities while other methods suffer significant degradation. Specifically, DAMA achieves an average WER of 58.80\% in the 1 hour setting. This score surpasses the fine-tuning baseline, which reaches 61.79\%. This suggests that updating all parameters in the low data regime leads to severe overfitting. The robustness of DAMA becomes evident when comparing it to methods like AdaLoRA and LoRA-XS. These baselines fail to adapt effectively, with error rates spiking above 72\%. While AdaLoRA employs dynamic rank allocation, it relies on data driven sensitivity scores to prune parameters. In extremely low resource settings, these scores become unreliable due to data sparsity, leading to poor architectural decisions. In contrast, DAMA uses a structural rank schedule based on the hierarchical nature of speech. This fixed prior guides the adaptation process without overfitting to the limited data. Consequently, DAMA consistently delivers high accuracy comparable to robust methods like LoRA and DoRA, proving its reliability even under challenging training conditions.


\subsection{Computational Efficiency}
\begin{figure}[t!]
\centering  
\includegraphics[width=0.9\linewidth]{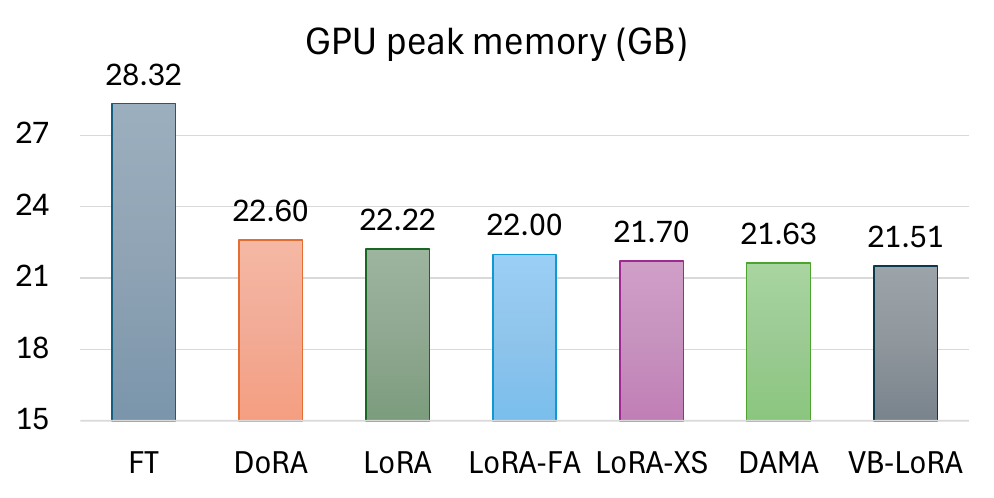}
\vspace{-2mm}
\caption{Average
Peak GPU memory usage for different adaptation methods across all 10 languages on Common Voice.}
\label{fg2}
\vspace{-5mm}
\end{figure}

 
\begin{figure}[t!]
\centering  
\includegraphics[width=\linewidth]{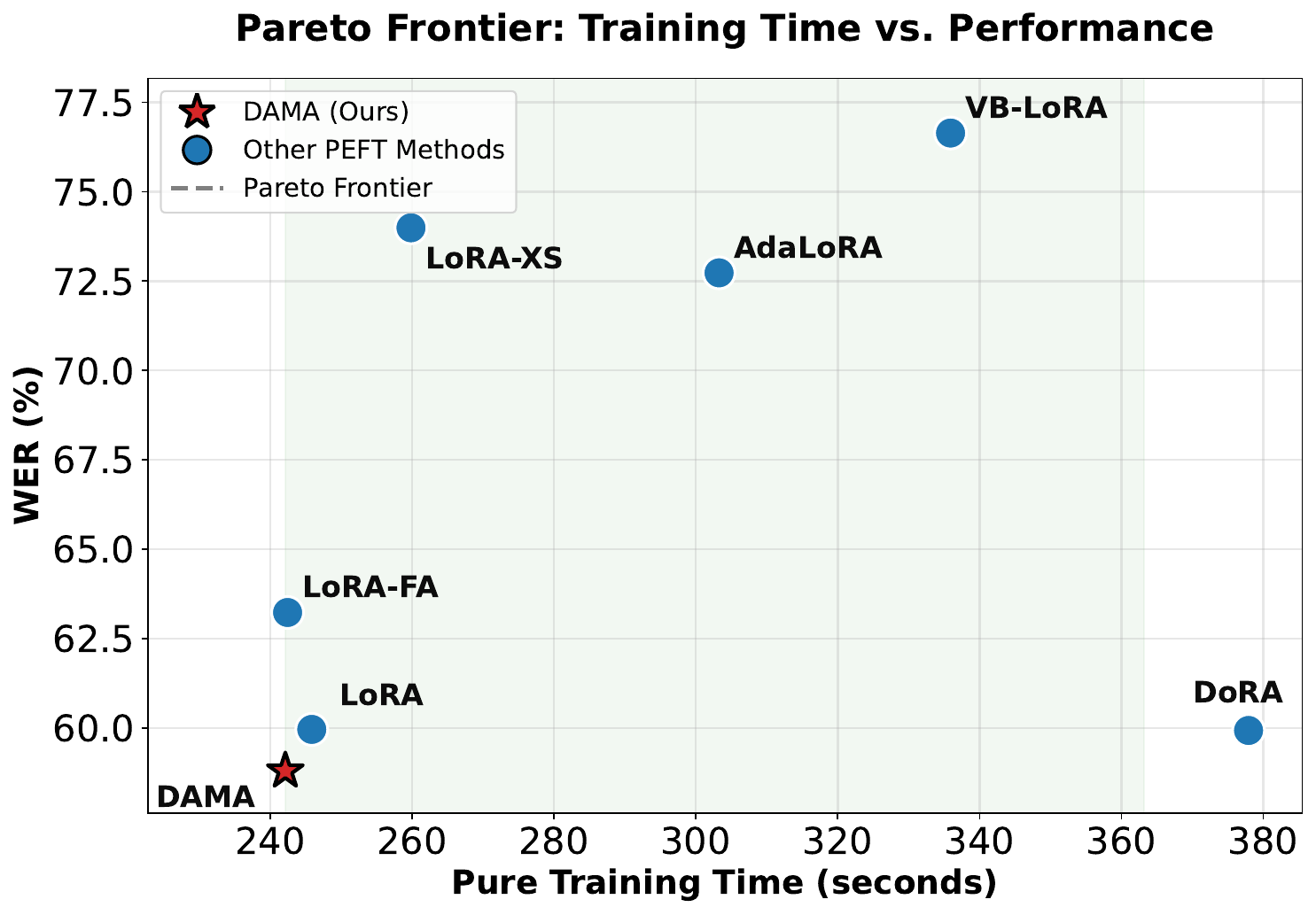}
\vspace{-8mm}
\caption{Pareto Frontier analysis of one-epoch adaptation time versus overall average WER.}
\label{fg4}
\vspace{-7mm}
\end{figure}

\paragraph{MACs.} 
We further evaluate the computational overhead using Extra MACs. As shown in Table 1, DAMA achieves the lowest overhead among all methods, requiring only 22.3G MACs. In strong contrast, standard LoRA demands 102.2G MACs, which is nearly five times higher. Furthermore, advanced methods like DoRA and VB-LoRA incur a massive computational cost, reaching 1415.4G and 758.3G MACs, respectively. 
While LoRA-FA and AdaLoRA achieve lower MACs compared to standard baselines, their MACs remain higher than ours. Additionally, both methods exhibit worse WER, especially on unseen languages.
This confirms that our sparse, protected basis design is the most suitable candidate for real-time applications.


\paragraph{GPU Peak Memory. }Figure~\ref{fg2} illustrates the peak GPU memory usage across different adaptation strategies. Compared to the resource-intensive fine-tuning, our approach significantly reduces the GPU peak memory usage from 28.32 GB to 21.63 GB, a substantial reduce of approximately 24\%, which makes our method much more accessible for resource-constrained settings.

When compared to other parameter-efficient methods, DAMA continues to demonstrate superior efficiency. It requires less memory than both standard LoRA and DoRA. 
This efficiency stems from our adaptive ranking strategy, which avoids unnecessary gradient updates. While VB LoRA achieves a comparable 
memory usage of 21.51 GB, it comes at a severe cost to accuracy with significantly worse WER. Consequently, DAMA offers the most favorable balance between memory cost and performance across all evaluated methods.

\paragraph{Training-time Efficiency Analysis.}
We further compare the adaptation time of a single epoch as shown in Figure 4. The 
Pareto figure shows the training time in seconds on the horizontal axis and WER on the vertical axis. DAMA occupies the optimal position at the bottom left corner, indicating it is the most efficient method among all baselines.

Specifically, DAMA achieves the fastest training time of 242.11 seconds while simultaneously maintaining the lowest WER of 58.80\%. In contrast, while LoRA FA matches our speed with 242.43 seconds, it suffers from a significantly higher WER of 63.23\%. This shows that simplifications made by LoRA FA compromise the model quality. Conversely, DoRA achieves a competitive error rate of 59.93\% but requires a much longer training time of 377.93 seconds. This 56\% increase in time makes DoRA less suitable for rapid adaptation. Finally, methods like AdaLoRA and VB-LoRA fall far behind the frontier, as they exhibit both slower training speeds and higher error rates. Therefore, DAMA offers the best balance of speed and accuracy for real time applications.



\subsection{Ablation Study}
\begin{table}[t]
\centering
\small
\setlength{\tabcolsep}{6pt}
\renewcommand{\arraystretch}{1}
\caption{Ablation study on the three key components of DAMA using the Common Voice dataset.}
\label{tab:abla_dama}
\resizebox{\linewidth}{!}{%
\begin{tabular}{c| c| c| c}
\toprule
Depth-Aware (U-shaped) & Basis-Protected (BPP) & SVD-based Initialization & Avg WER \\
\midrule
Uniform            & \cmark & \cmark & 43.67 \\
\cmark & Full Adaptation    & \cmark & \textbf{42.98} \\
\cmark & \cmark & Random    & 43.95 \\
\bottomrule
\end{tabular}}
\vspace{-5mm}
\end{table}
We investigate the contribution of the three designs in DAMA: the Depth-Aware Rank Schedule, the SVD-based Initialization and BSP in Table 3. First, we replace our depth-aware schedule with a uniform rank distribution, which increases the WER to 43.67\% even with more complex adapter. This result confirms the ``U-shaped'' plasticity hypothesis: the model requires high adaptability at the early and late layers, but strictly limited interventions in the middle layers to preserve performance.

Second, we compare our SVD-based initialization method against a standard random initialization. The random approach yields a higher error rate of 43.95\%, proving that SVD constrains LoRA updates to directions that
minimally impact the language-agnostic semantic
subspace and maintains the u-shape for improved performance. We finally replace the proposed BSP with full adaptation, yielding a negligible accuracy gain of only 0.22\% (43.20\% vs. 42.98\%). However, this minor gain comes at the cost of structural safety and more parameters in the middle layers. 

\section{Conclusions}

This work presents the first systematic investigation of layer-wise plasticity in MASR and reveals a distinctive U-shaped pattern. Building on this finding, we introduce DAMA, a novel adaptation framework that leverages this U-shaped structure for more efficient and effective adaptation to new languages. Experimental results show that DAMA achieves strong performance across languages, especially in low-resource settings, while significantly reducing the number of trainable parameters and improving efficiency in memory usage, training time, and floating-point operations. These results highlight the importance of tailoring adaptation strategies to the internal organization of foundation models. Our findings open new avenues for developing scalable and robust multilingual and low-resource adaptation methods by harnessing model-internal representations, paving the way for more accessible and efficient speech technologies.


\section*{Limitations}
While DAMA demonstrates strong robustness, there are several avenues for future improvement. First, although our approach has been validated on 18 linguistically diverse languages, expanding evaluation to an even broader range, including extremely rare dialects, will further test the generalizability of the "U-shaped" prior. Second, our method is specifically optimized for low-resource adaptation by ensuring structural integrity; in high-resource scenarios, relaxing the constraints on the Semantic Valley may unlock even greater performance gains. Finally, extending this work to explore whether the Semantic Valley phenomenon supports other downstream tasks beyond ASR presents an exciting direction for future research.

\section*{Ethics Statement}
All the data used in this paper are publicly available and are used under the following licenses: the Creative Commons BY-NC-ND 4.0 License and Creative Commons Attribution 4.0 International License, the TED Terms of Use, the YouTube’s Terms of Service, and the BBC’s Terms of Use. 


\bibliography{main}

\clearpage

\appendix
\section{Details of the Depth-aware Analysis}
\subsection{Probing Methodology}

To ensure that our analysis reflects the intrinsic representations of the model rather than the capacity of the probe, we employed a lightweight linear probing protocol. For probing, we keep the Whisper backbone completely frozen and only train a lightweight linear classifier on top of its hidden states. For a chosen layer \(l\) in decoder, we extract the full sequence of hidden states for each utterance, then apply a mean temporal pooling strategy to obtain a single dimensional representation per example. These pooled representations are passed to a single linear layer that predicts the language ID, with locales mapped to integer labels via a normalized locale vocabulary. We optimize only this linear probe using AdamW with a learning rate of \(1\times10^{-3}\), cross-entropy loss for language identification, with batch size 256 and 5 epochs in our experiments, selecting the best model by validation loss before reporting final accuracy on the test set.

\subsection{Analyzed Languages}

To validate the universality of the U-shaped plasticity profile, we selected a diverse set of languages from the Common Voice dataset.

\begin{itemize}
    \item \textbf{Seen Languages:} English, Turkish, Russian, German, and Chinese.
    \item \textbf{Unseen Languages:}
    Kinyarwanda, Esperanto, Kabyle, Luganda, Meadow Mari, Central Kurdish, Abkhaz, Kurmanji Kurdish, Frisian, and Interlingua.
\end{itemize}

\subsection{Linguistic Rationale for Language Selection}

To prevent selection bias, , we selected the "Seen" languages to maximize typological diversity. We specifically chose languages that diverge from English across three structural dimensions, ensuring that the observed ``Semantic Valley'' is a universal phenomenon.

\begin{itemize}
    \item Syntactic Order: We included SOV languages (e.g., Turkish) to contrast with the SVO structure of English, testing the model's robustness to significant word-order shifts.
    \item Morphology: We selected fusional languages (e.g., Russian, German) to challenge the plastic early layers with complex inflections and token sparsity.
    \item Information Density: We included Chinese to validate the semantic valley's stability even when processing high-density characters that differ radically from alphabetic subwords.
\end{itemize}

Consequently, for the "Unseen" target languages, we adhered to the established CL-MASR benchmark~\citep{cl-masr} (e.g., Kinyarwanda, Luganda) to ensure fair, reproducible comparisons with baselines.

\section{Experiments}
\subsection{ Datasets and Language Statistics }
In this subsection, we introduce how we design the dataset for experiment. For the Common Voice `Unseen' languages, we adopted the standard CL-MASR benchmark~\citep{cl-masr} including languages such as Kinyarwanda and Luganda—to guarantee fair and reproducible baseline comparisons.

For the FLEURS dataset. We selected the Whisper Seen-Weak set (Hindi, Welsh, Belarusian, Persian, Swahili) by looking for languages where WER drops clearly from Whisper-tiny to larger checkpoints, but where there is still visible room for improvement. This pattern suggests that scaling helps, so these languages are suitable for PEFT because they are not ``stuck'' at an extreme error level even at larger models. We also selected by geographic diversity under the FLEURS grouping and diversity in writing systems and linguistic structure, so that any gains we see are less likely to be specific to one region or one script.

The priority was to cover multiple FLEURS regions and to include both a near-neighbor transfer case (Asturian) and structurally diverse languages: Luganda as a Bantu language from Sub-Saharan Africa, Central Kurdish from the Middle East with morphologically rich structure and Perso-Arabic script, Oriya with Odia script from South Asia, and Cebuano as an Austronesian language from South-East Asia, so improvements are informative across different linguistic and geographic conditions.

\begin{table}[t]
\centering
\small
\resizebox{!}{0.55\linewidth}{
\begin{tabular}{llrrr}
\toprule
\textbf{Language} & \textbf{ISO 639-1} & \multicolumn{3}{c}{\textbf{Duration (minutes)}} \\
\cmidrule(lr){3-5}
 &  & \textbf{Training} & \textbf{Validation} & \textbf{Test} \\
\midrule
\multicolumn{5}{c}{\textbf{Common Voice}} \\
\midrule
Kinyarwanda      & rw    & 600 & 60 & 60 \\
Esperanto        & eo    & 600 & 60 & 60 \\
Kabyle           & kab   & 600 & 60 & 60 \\
Luganda          & lg    & 600 & 60 & 60 \\
Meadow Mari      & mhr   & 600 & 60 & 60 \\
Central Kurdish  & ckb   & 484 & 60 & 60 \\
Abkhaz           & ab    & 600 & 60 & 60 \\
Kurmanji Kurdish & kmr   & 296 & 60 & 60 \\
Frisian          & fy-NL & 330 & 60 & 60 \\
Interlingua      & ia    & 313 & 60 & 60 \\
\midrule
\multicolumn{5}{c}{\textbf{FLEURS}} \\
\midrule
Hindi      & hi    & 399 & 60 & 60 \\
Welsh      & cy & 600 & 60 & 60 \\
Belarusian       & be    & 571 & 60 & 60 \\
Persian      & fa    & 600 & 60 & 60 \\
Swahili      & sw    & 600 & 60 & 60 \\
Luganda       & lg    & 600 & 60 & 60 \\
Central Kurdish   & ckb    & 584 & 60 & 60 \\
Asturian     & ast    & 452 & 60 & 60 \\
Cebuano      & ceb    & 600 & 60 & 60 \\
Oriya       & or    & 206 & 60 & 60 \\
\bottomrule
\end{tabular}}
\caption{Data duration (minutes) per language split into training, validation, and test sets.}
\label{tab:data_duration_by_language}
\end{table}

\begin{table*}[!t]
\centering
\small
\caption{Results of the Common Voice dataset by language for different adaptation methods.}
\vspace{-3mm}
\resizebox{!}{0.15\linewidth}{
\begin{tabular}{lcccccccc}
\toprule
\textbf{Lang} & \textbf{FT} & \textbf{LoRA} & \textbf{DoRA} & \textbf{LoRA-FA} & \textbf{LoRA-XS} & \textbf{VB-LoRA} & \textbf{AdaLoRA} & \textbf{DAMA} \\
\midrule
ab    & 60.12 & 61.59 & 61.79 & 64.58 & 81.09 & 73.91 & 73.35 & 62.01 \\
ckb   & 51.42 & 50.64 & 51.04 & 51.04 & 63.81 & 58.77 & 59.03 & 50.97 \\
eo    & 19.64 & 15.36 & 15.51 & 16.11 & 20.92 & 22.36 & 18.61 & 15.36 \\
fy-NL & 28.92 & 26.45 & 25.63 & 29.87 & 40.94 & 36.17 & 34.27 & 26.41 \\
ia    & 16.49 &  9.04 &  9.31 & 10.25 & 13.60 & 12.74 & 11.52 &  9.33 \\
kab   & 63.73 & 68.29 & 68.75 & 72.91 & 85.34 & 79.29 & 79.71 & 67.14 \\
kmr   & 39.97 & 39.40 & 39.52 & 42.19 & 55.13 & 45.36 & 48.97 & 38.61 \\
lg    & 58.37 & 59.42 & 60.51 & 63.81 & 77.84 & 68.67 & 69.59 & 60.07 \\
mhr   & 32.26 & 33.06 & 33.19 & 36.56 & 50.32 & 44.67 & 43.33 & 32.93 \\
rw    & 67.81 & 69.24 & 70.89 & 72.66 & 84.65 & 77.48 & 78.19 & 69.13 \\
\bottomrule
\end{tabular}}
\vspace{-2mm}
\label{tab:multilingual_asr_results}
\end{table*}

\begin{table*}[t!]
\centering
\small
\caption{Results of the FLEURS dataset by language for different adaptation methods.`*' means seen languages.}
\vspace{-3mm}
\resizebox{!}{0.15\linewidth}{
\begin{tabular}{lcccccccc}
\toprule
\textbf{Lang} & \textbf{FT} & \textbf{LoRA} & \textbf{DoRA} & \textbf{LoRA-FA} & \textbf{LoRA-XS} & \textbf{VB-LoRA} & \textbf{AdaLoRA} & \textbf{DAMA} \\
\midrule
ast & 22.69 & 17.28 & 17.32 & 19.17 & 24.56 & 19.93 & 21.06 & 17.64 \\
be*  & 23.23 & 19.43 & 19.53 & 20.03 & 27.91 & 22.88 & 22.78 & 18.69 \\
ceb & 21.45 & 18.91 & 17.80 & 19.52 & 23.37 & 20.67 & 20.16 & 18.70 \\
ckb & 57.12 & 58.29 & 58.18 & 60.47 & 71.79 & 66.43 & 66.76 & 58.83 \\
cy*  & 28.98 & 27.73 & 27.52 & 27.77 & 28.96 & 27.90 & 29.47 & 27.77 \\
fa*  & 23.16 & 19.95 & 19.91 & 20.05 & 22.87 & 22.30 & 21.53 & 20.15 \\
hi*  & 25.64 & 24.78 & 24.78 & 24.90 & 27.38 & 26.15 & 26.90 & 24.82 \\
lg  & 58.81 & 62.95 & 60.72 & 63.85 & 72.21 & 64.92 & 68.97 & 61.71 \\
or  & 90.18 & 78.22 & 78.28 & 79.77 & 84.22 & 81.78 & 84.89 & 79.37 \\
sw*  & 36.76 & 34.08 & 34.46 & 35.50 & 44.37 & 40.79 & 41.35 & 34.88 \\
\bottomrule
\end{tabular} }
\vspace{-2mm}
\label{tab:multilingual_asr_results_2}
\end{table*}

\subsection{Baseline methods}
To comprehensively validate the proposed DAMA framework, we compared it against three categories of adaptation methods. This selection covers the spectrum from standard uniform approaches to advanced dynamic allocation strategies.

\paragraph{1. Uniform Adaptation Baselines:}
These methods apply a fixed rank across all layers, treating the model as a homogeneous structure.
\begin{itemize}
    \item \textbf{LoRA~\citep{hu2022lora}:} The most widely used PEFT method. It injects trainable low-rank matrices ($A$ and $B$) into every layer with a uniform rank.
    \item \textbf{DoRA~\citep{dora}:} A robust variant of LoRA that decomposes weights into magnitude and direction. It serves as a strong baseline for accuracy but suffers from high computational cost during training.
\end{itemize}

\paragraph{2. Data-Driven Dynamic Baselines:} These methods attempt to allocate parameters based on data sensitivity, which contrasts with our structure-driven approach.
\begin{itemize}
    \item \textbf{AdaLoRA~\citep{zhangadaptive}:} It dynamically allocates the rank budget among layers based on the importance scores derived from gradients. As discussed in Section 6.2, this method often struggles in low-resource settings where gradient signals are noisy.
\end{itemize}

\paragraph{3. Efficiency-Focused Variants:}
These methods prioritize parameter or memory efficiency.
\begin{itemize}
    \item \textbf{LoRA-FA~\citep{lorafa}:} Freezes the projection-down matrix $A$ (randomly initialized) and only trains $B$. While memory-efficient, it lacks the structural initialization of our Basis-Protected Projection.
    \item \textbf{LoRA-XS~\citep{loraxs}:} Utilizes Singular Value Decomposition (SVD) to perform static compression of the weight updates.
    \item \textbf{VB-LoRA~\citep{li2024vb}:} Employs a shared ``Vector Bank"" to construct low-rank matrices. This represents an extreme compression approach but often compromises performance on unseen languages.
\end{itemize}

Comparing DAMA against these diverse baselines allows us to verify that our  strategy outperforms both uniform methods and data-driven methods (which overfit in low-resource regimes).

\subsection{Baseline methods hyperparameter setting}

\paragraph{Common Training Settings:} 
All methods use a training batch size of 6, a validation batch size is 16, and a sample rate is 16000 Hz.The maximum target sequence length is 448. All methods train for 2 epochs with AdamW optimizer, maximum gradient norm of 5.0, FP16 precision, and learning rate scheduling via NewBobScheduler (improvement threshold 0.0025, annealing factor 0.8). For the Common Voice dataset, utterances longer than 10 seconds are filtered, the maximum generation tokens is 80. For the FLEURS dataset, utterances longer than 30 seconds are filtered, the maximum generation tokens is 120. 

\paragraph{Uniform Low-Rank Methods}
For uniform low-rank methods, we maintained consistent structural constraints across all layers. Specifically, both LoRA and DoRA were configured with a fixed rank of $r=64$ and alpha $\alpha=64$. The primary distinction lies in their optimization approach: DoRA applies weight decomposition without dropout to isolate magnitude updates, whereas LoRA utilizes a dropout rate of 0.1. 

\paragraph{Advanced \& Efficiency Variants}
Regarding advanced efficiency variants, we adopted specific configurations tailored to their dynamic architectures. AdaLoRA utilized an adaptive budget allocation strategy, initializing with a rank of 48 and gradually pruning to a target rank of 32 based on sensitivity scores smoothed by an EMA factor of 0.85. In contrast, VB-LoRA employed an extreme compression approach using a vector bank of 90 vectors (dimension 1280) with Top-2 selection; notably, this required a higher learning rate of $1e^{-3}$ to ensure convergence within the restricted parameter space. Finally, LoRA-XS was constructed via SVD decomposition (10 iterations) of a pre-trained LoRA module ($r=64$), where we froze the resulting low-rank matrices and only trained the intermediate $r \times r$ latent mapping matrix. We also evaluated LoRA-FA, which initializes the projection-down matrix $A$ randomly and freezes it, exclusively updating the projection-up matrix $B$. This variant similarly employed rank $r=64$, serving as a baseline for static subspace constraints.

\section{Additional results}

\subsection{Results by different languages.}
The detailed performance on the Common Voice and FLEURS datasets is presented in Table~\ref{tab:multilingual_asr_results} and Table~\ref{tab:multilingual_asr_results_2}.
This evaluation encompasses a diverse set of low-resource languages, categorized into ``Unseen'' targets and ``Seen-Weak'' languages (with `*' in the table).
Specifically, these results highlight the superior generalization capability of DAMA compared to uniform baselines like LoRA and data-driven variants like AdaLoRA.
While standard methods suffer from degradation on unseen languages due to overfitting, DAMA consistently maintains high accuracy across both categories.

\subsection{Forgetting estimation}

\begin{table}[t]
\centering
\small
\caption{Comparison of Catastrophic Forgetting on English (Seen) after adapting to Kinyarwanda (Unseen).}
\resizebox{!}{0.09\linewidth}{
\begin{tabular}{lrrrrr}
\toprule
\textbf{Lang} & \textbf{Base} & \textbf{FT} & \textbf{LoRA} & \textbf{DoRA} & \textbf{DAMA} \\
\midrule
English & 11.09 & 105.50 & 12.93 & 13.09 & 12.56 \\
Kinyarwanda & --    &  67.81 & 69.24 & 70.89 & 69.13 \\
\bottomrule
\end{tabular}}
\vspace{-5mm}
\label{tab:forgetting}
\end{table}

To evaluate whether the adaptation process damages the model's existing knowledge, we analyzed the performance on the source language (English) after training on a target low-resource language (Kinyarwanda). The results are presented in Table~\ref{tab:forgetting}.
Specifically, the data reveals the severe risk of Catastrophic Forgetting associated with unconstrained updates.
As shown in the table, Fine-Tuning (FT) causes the WER on English to spike dramatically from the baseline 11.09\% to 105.5\%.
This indicates that while FT adapts to the new language, it completely overwrites the model's pre-trained semantic core.

In stark contrast, DAMA demonstrates superior stability then .
It maintains an English WER of 12.56\%, which is significantly closer to the original baseline and outperforms both LoRA (12.93\%) and DoRA (13.09\%).
Consequently, this confirms that our Basis-Protected Projection (BPP) successfully locks the "Semantic Valley," ensuring that the model learns new languages without sacrificing its intrinsic capabilities.

\end{document}